\documentclass[12pt,preprint]{aastex}

\usepackage{epsf}
\usepackage{graphics}

\newcommand{\kms}{km\,s$^{-1}$}
\newenvironment{inlinefigure}{
\def\@captype{figure}
\noindent\begin{minipage}{0.999\linewidth}\begin{center}}
{\end{center}\end{minipage}\smallskip}

\slugcomment{Accepted to the Astrophysical Journal}
\shorttitle{Globular Clusters in VCC~1386}
\shortauthors{Conselice}

\def\kms{km s$^{-1}$}

\begin{document}

\title{Keck Spectroscopy of Globular Clusters in the Virgo Cluster Dwarf Elliptical VCC~1386}
 
\author{Christopher J. Conselice$^{1,2}$}
 
\altaffiltext{1}{California Institute of Technology, Mail Code 105-24, Pasadena
CA 91125}
\altaffiltext{2}{National Science Foundation Astronomy \& Astrophysics Fellow}

\begin{abstract}

We present the results of a Keck spectroscopic study of globular clusters
associated with the Virgo Cluster dwarf elliptical VCC~1386. We
analyze blue spectroscopic absorption lines from 3500-5500 \AA\ for 13
globular cluster candidates and 
confirm that five are associated with VCC~1386.     By comparing metal and  
Balmer line indices of these
globular clusters with $\alpha$-enhanced single stellar population
models we find that these systems
are metal poor with [Fe/H] $< -1.35$ and old, with ages $> 5$ Gyr at
3 $\sigma$ confidence, placing
their formation at $z \gtrsim 1$. This is one of the first 
spectroscopic studies of globular clusters surrounding dwarfs in
a cluster, revealing that some low mass galaxies in rich environments
form at least part of their stellar mass early in the history of the universe.
We further find that the luminosity weighted stellar
population of VCC~1386 itself is younger, and more metal rich than its
globular clusters, consistent with $($V$-$I$)_{0}$ colors from Hubble
Space Telescope imaging.   This implies that VCC~1386, like Local Group
dEs, has had multiple episodes of star formation.  Globular clusters 
associated with low luminosity systems however appear to be roughly as old as
those associated with giant galaxies, contrary to the
`downsizing' formation of their bulk stellar populations.

\end{abstract}
 
\section{Introduction}

Galaxy formation is thought by many to be a process whereby massive 
galaxies are built from smaller ones.  This hierarchical process is
now well modeled and is a natural outcome of a lambda dominated universe 
with cold dark matter (e.g., Cole et al. 2000).  
Predictions based on simulations that incorporate these assumptions
compare well with the large scale features of 
the universe measured with microwave background experiments
(e.g., Spergel et al. 2003), and large scale galaxy surveys 
(Tegmark et al. 2004).   However,  there are several problems with 
the basic hierarchical model, mostly on the 
level of galaxies, and especially when predicting
properties of the lowest mass systems 
(e.g., Moore et al. 1999).  This includes predicting
too many low mass companions around
galaxies like our Milky Way.  These models also predict  too 
few major mergers at $z > 2$ (Conselice 2005), and have a difficult
time reproducing observations of all dwarfs in rich galaxy clusters
(e.g., Conselice et al. 2003a).    Currently, it is assumed that
differences between models and observations are due to an
incomplete understanding of
the physics behind star formation, although problems
may also exist on a deeper more fundamental level as well.

Other tests of the hierarchical galaxy formation model are therefore
needed to fully assess its assumptions and predictions.  
A potentially powerful way to do this is through examining low-mass galaxies, 
which are predicted to be, on average,
the first galaxies in the universe.    If the hierarchical 
model is correct these low-mass galaxies are possibly survivors of the merger 
process, and
should contain some of the oldest stellar populations.
Theory also predicts that the first galaxies
should form in the densest environments (e.g., Springel et al. 2001; 
Tully et al. 2002), and thus cluster dwarf ellipticals potentially host
the first stars that formed in the universe, perhaps before
reionization (e.g., Bullock et al. 2000).  This is particularly the case
for dwarfs that exist near the center of clusters, where the oldest
stars in the nearby universe are predicted to be found (White \& Spergel 2000).

There is considerable evidence however that dwarf ellipticals, and lower
mass galaxies in general, formed or entered clusters after the 
giant galaxies.  For example,
measurements of the integrated light in dwarf ellipticals reveal 
stellar population ages that are younger than the
stars that make up giant ellipticals (e.g., Poggianti et al. 2001; 
Caldwell, Rose \& Concannon 2003; Rakos \& Schombert 2004).  Likewise,
the faint end of the red sequence in clusters is not formed until $z < 1$
(De Lucia et al. 2004).   This implies that the luminosity
weighted stellar populations of lower mass galaxies are younger than
the stars in giant ellipticals.   This would appear to be an
`anti-hierarchical' method of galaxy formation, such that the lowest mass
systems formed after the more massive ones.  This has been
seen in field galaxies as well and has been described as `downsizing'
(Cowie et al. 1996).   

Curiously, it also appears that a significant
fraction of faint dEs with M$_{\rm B} > -15$ originate from a process
different from that which formed the brighter dEs and Es.  Evidence for this
includes dE radial velocity distributions similar
to infalling spirals (Conselice et al. 2001), and dEs with integrated light
that appear young and/or metal rich (Conselice et al. 2003a; 
Rakos \& Schombert 2004).  These dEs are potentially produced from higher
mass galaxies that 
became stripped of mass after infalling into 
clusters (e.g., Conselice 2002).  However,
the population of nucleated cluster dwarf ellipticals
tend to have properties suggesting they are the low 
luminosity counterparts of giant ellipticals, such as a high
globular cluster specific frequency (Miller
et al. 1998).  Some dEs also fall along the color-magnitude relationship,
while others do not (e.g., Rakos et al. 2001; Conselice et al. 2002, 2003a).  
It is thus not yet clear when, or how, low mass galaxies form within a 
dense environment, or their relationship to dwarf galaxies in the Local
Group and giant cluster ellipticals.

We address these issues by studying the stellar population
properties of globular clusters surrounding a dwarf elliptical, VCC~1386,
in  the Virgo cluster.   Globular clusters 
are ideal targets for this type of analysis as they can be studied
as single stellar populations, and may retain clues to the earliest galaxy 
formation mechanisms (e.g., Forbes et al. 2004). Our target,
VCC~1386, is a M$_{\rm V}$ = $-16.25$ ($\sim$ L$_{*}$+5) 
dE with a system of globular cluster candidates, 13 of which we obtain 
spectroscopy for using the Keck I telescope.   We confirm that five globular 
clusters are associated 
with VCC~1386, while three objects are foreground stars, with 
the remainder having too faint a signal  to place constraints on 
membership.  Our main conclusion is that some of
the globular clusters surrounding 
VCC~1386 contain old and metal poor stars, in contrast to the dwarf 
itself, which contains a younger and more metal rich luminosity weighted 
stellar population. This suggests that VCC~1386 contains
multiple star formation episodes, and some globulars in this system
are as old as, or older than, those in giant ellipticals.
Throughout this paper we use a distance to the
Virgo cluster of 16.7 Mpc (Whitmore et al. 1995).

\section{Observations}

We selected  VCC~1386 globular cluster candidates  from WFPC2 Hubble
Space Telescope (HST) imaging
in the F555W (V) and
F814W (I) bands. This imaging is from an HST dwarf elliptical
galaxy snap-shot survey of the Virgo and Fornax clusters
(e.g., Lotz et al. 2004).  Exposure times were 2 $\times$ 230
seconds in the F555W band,  and 300 seconds in the F814W band.  
For the VCC~1386
system, candidates globular clusters were selected based on having
(V$-$I)$_{0}$ colors consistent with previous observations of
globular clusters, with (V$-$I)$_{0}$ $< 1.3$.   
We measured magnitudes in the V and I bands
for each globular cluster candidate using a 2\arcsec\, aperture with APHOT
on IRAF, and corrected for galactic extinction using the reddening corrections
in Schlegel et al. (1998).   Based on their colors, we obtained a 
sample of 28 globular cluster candidates.

The spectroscopic observations we present were taken with the 
Keck Low Resolution Imaging Spectrometer
(LRIS; Oke et al. 1995) on the Keck I telescope in February 2004. 
Using one LRIS mask, we obtained
spectroscopy for 13 globular cluster candidates.  These clusters were
located from near the center
of VCC~1386, to a projected radius of 3 kpc. This system is
shown in Figure~1 with the inner globular candidates with spectroscopy
circled.  This spectroscopy was acquired under good (seeing
$\sim 0.6$\arcsec), but
unphotometric conditions with a total exposure time lasting $\sim$ 7 hours. 
We obtained spectroscopy with both the
red and blue sides of LRIS (see Steidel et al. 2004 for a description
of LRIS-B). On the blue side we utilized a 400 lines/mm grism blazed
at 3400 \AA, while on
the red side we used the 600 lines/mm grating blazed at 5000 \AA. This produced
an effective resolution of 8 \AA\, and 6 \AA, respectively.  The wavelength 
range probe
was $\sim 3500-5500$ \AA\, for LRIS-B, with wavelength calibration performed
using HgKrXeNe comparison arc lamp spectra.
The spectra was bias subtracted, flat-fielded and rectified using 
IRAF reduction techniques for multi-slit data, combining the various frames 
into a single one, and extracting the spectra for the individual globular 
clusters from this combination. 

\section{Stellar Populations of the VCC~1386 Globular Clusters}

Using our spectra we identified features
such as the Calcium H+K lines, Fe lines, as well as Balmer lines, such
as H$\beta$, in our spectra. 
We derived from these lines the radial velocities of the globular clusters
through cross correlating the measured absorption lines with a known early type
galaxy spectrum and by identifying known lines by eye.   Five of these 
systems (\#1,2,3,4,6) (Table~1) are at the radial velocity of the host 
galaxy, VCC~1386, at
$\sim$1300 \kms\, (Simien \& Prugniel 2002).   Three globular cluster candidates 
have radial velocities that place them within our own galaxy, and two of
these are at a large projected distance  from the center
of VCC~1386.  All of the confirmed globular clusters are within
1.5 kpc of the center of VCC~1386. 

\subsection{Metallicity and Age}

We place constraints on the stellar populations that make up our globular
cluster sample by utilizing the spectra and colors 
of the five confirmed members. 
We use Lick line indices, measured from the spectra,
to determine metallicities and
ages of the globular clusters (Worthey 1994; Worthey \& Ottaviani 1997; Trager
et al. 2000; Thomas et al. 2003), although not
every absorption feature could be measured for every globular cluster. 
We measured the equivalent widths of the Lick indices after
smoothing our spectra by a wavelength dependent 
Gaussian kernel to match the Lick ID resolution 
of $\sim 9$\AA\, (Worthey \& Ottaviani 1997).  Our data is not flux
calibrated, but flux calibrated indices are $< 1$\% different from
non-flux calibrated ones (e.g., Strader et al. 2003), and thus not likely contributing
significantly to our error budget. 
We measured in our analysis the Balmer lines H$\beta$, H$\delta_{\rm F}$, 
H$\gamma_{\rm F}$, and the metal lines
Fe 5270, Fe 5335 and Mg $b$ using the updated passband and continuum 
wavelengths from Worthey (1994) and Worthey \& Ottaviani (1997).  
Because we were unable to obtain Lick
standard stars during our observations, we are not able to produce a direct 
offset between our indices and the Lick system. However using the same setup
with LRIS under identical conditions  these offsets are found to be
smaller than the observational errors, and are not often applied
(Brodie et al. 2005).
 
We determine the ages and metallicities of the stellar populations in our
globular clusters through comparisons to Bruzual and Charlot (2003) and
Thomas et al. (2003) single stellar population models.  First, to constrain
the ages of our globular clusters, we compare our measured Balmer indices
(H$\beta$, H$\delta_{\rm F}$, H$\gamma_{\rm F}$) with
Bruzual \& Charlot (2003) models (Figure~2).  The three
horizontal lines in Figure~2 show the measure values of the H$\beta$, 
H$\delta_{\rm F}$, and H$\gamma_{\rm F}$ indices.  The four lines 
over-plotted show the modeled evolution of these indices with time.  
While several ages are possible for the globular clusters
based on these indices, they are generally consistent with being older than 
5 Gyr, unless they have metallicities higher than [Fe/H] = $-0.64$, at
a confidence $> 3 \sigma$. 

We cannot make definite conclusions regarding the ages of the globular
clusters based solely on Balmer indices.  However, when we compare Bruzual
\& Charlot (2003) models of (V$-$I) color and the H$\beta$ index, we find that
old ages are preferred (Figure~3).  Figure~3 shows the modeled
evolution of (V$-$I) colors as a function of
H$\beta$ at two different metallicities, [Fe/H] = $-1.64$ 
and [Fe/H] = $-0.64$.  The evolution is such that that 
the typical globular cluster H$\beta$ value of 2.5 \AA\, 
is reached several times during the evolution of these single stellar
populations 
(SSP).  However, after about 1 Gyr, the colors of these SSP models are redder 
than (V$-$I) $\sim 0.6$.
This effectively limits the possibility that these globular clusters are 
extremely young systems,
and is consistent with their being generally older than 
5 Gyrs.

Constraining metallicity is also not straightforward, as most metal lines 
such as Mg $b$
are influenced by the ages of stellar populations as well as by their
metallicity.   However by combining
a metal index with a Balmer index we can constrain the luminosity
weighted ages and 
metallicities of stellar populations. 
As such, we compare our measured indices to the $\alpha$-enhanced 
SSP models of Thomas et al. (2003), using the $\alpha$ = 0.3 models for 
the three globulars
(\#2,4,6) with a high enough signal to noise to accurately measure these indices.
We utilize the index $<$Fe$>$ = 0.5 $\times$ (Fe5270 + Fe5335), the average of
the Fe5270 and Fe5335 indices, as a measure of metallicity when compared
with H$\beta$.  We plot these values for our globular clusters on Figure~4a. 
The corresponding H$\beta$ vs. $<$Fe$>$ point
for the body of VCC~1386 is also plotted on Figure~4, where
H$\beta$ = 2.21$\pm$0.24, and $<$Fe$>$ = 2.15$\pm$0.23  (Geha et al. 2003). 

Figure~4 shows that VCC~1386's globular clusters are 
located in the low metallicity and old age part of the H$\beta$-$<$Fe$>$
diagram, with ages $\sim 8$ Gyr, and metallicities [Fe/H] $< -1.35$.  By
comparing with the Thomas et al. (2003) models, we can place statistical
limits on the ages of our stellar populations.  At 3 $\sigma$ confidence
all three globulars with measured H$\beta$ and $<$Fe$>$
indices are older than 5 Gyr
and have a metallicity [Fe/H] $< -0.33$.
When we compare the H$\beta$ and the [$\alpha$/Fe] insensitive index
[MgFe]' = [Mg $b$ $\times$ (0.72$\times$Fe5270+0.28$\times$Fe5335)]$^{1/2}$ 
(Thomas et al. 2003) we obtain the same constraints on ages and 
metallicities for
globular cluster \#2 (Figure 4b).  We also use the M31 globular cluster 
relation
between metallicity and (V$-$I)$_{0}$ color, [Fe/H] = 4.22$\pm$0.39 
(V$-$I)$_{0}$ $-$ 5.39$\pm$0.35 (Barmby et al. 2000) to place a photometric
limit on the metallicities of these systems.  
Using this relationship, we find that the globular cluster metallicities vary 
between [Fe/H] = $-1.3$ and
[Fe/H] = $-1.9$, consistent with their position on Figure~4. 

Interestingly, one globular cluster (\#4) has Balmer indices that are 
slightly larger than the other globular clusters.
This globular cluster is potentially either
more metal poor and as old as the others, or has a similar metallicity, but 
is younger.  This globular
is in fact located towards the center of VCC~1386 and therefore might have formed along with the
bulk of the stellar populations that make up the body of VCC~1386.  However, 
this globular cluster is not statistically
inconsistent with having an old ($> 8$ Gyr) age and a low metallicity
with [Fe/H] $ < -1.7$.  

Finally, in comparison to these globular clusters, the body of VCC~1386 
itself has a higher $<$Fe$>$ index at $\sim 5~\sigma$ confidence, but a 
similar H$\beta$ index, 
giving it a higher single stellar population metallicity and a younger 
age (Figure~4).  A comparison of the 
distribution of (V$-$I)$_0$
colors for our globular clusters and the light profile of the dE
itself (Figure~5) reveals that VCC~1386's globular clusters are more 
metal poor than the underlying light of VCC~1386, assuming that 
(V$-$I)$_{0}$ color is a tracer
of metallicity. The colors of the globular clusters
are in all but one case bluer than the underlying light at a given
projected distance from the center of VCC~1386.  If we
interpret these (V$-$I)$_{0}$ colors as a metallicity
indicator, then these globular clusters have 
a lower metallicity than the dE itself at every projected distance.  

\subsection{Formation Timescales with [$\alpha$/Fe] as a constraint}

Measuring the ratio of $\alpha$ elements to Fe, [$\alpha$/Fe] allows us to 
place constraints on the star formation timescales of our
globular clusters.  The reason is that Type II supernovae, which produce
$\alpha$ elements, occur over a much quicker time-scale than Type Ia SN,
which produce Iron.   By examining the location of our globular clusters in
a Mg $b$ vs. $<$Fe$>$ diagram, we can derive the likely values of
[$\alpha$/Fe] for our globular clusters (Figure~6).   The Thomas et al. (2003) models 
predict that at higher $\alpha$, the value of Mg $b$ is larger
at a given $<$Fe$>$.  The location of globular cluster \#2
is consistent with a high [$\alpha$/Fe] ratio.  This is not
an unambiguous result as the Thomas et al. (2003) models converge at low
metallicity, and our errors are not small enough to reject lower
[$\alpha$/Fe] ratios. In contrast, the average values of Mg $b$
and $<$Fe$>$ for VCC~1087's globular clusters place them in a low 
[$\alpha$/Fe] region of Figure~6.
If globular cluster \#2 in VCC~1386 indeed has a high
[$\alpha$/Fe] ratio it could be either
an indication that it formed in a rapid star formation event, or that 
the IMF of the star formation was top heavy, and more massive stars 
were available for producing $\alpha$ elements thorough Type II SN. 

\section{Interpretation \& Discussion}

Hierarchical structure formation models predict that 
within a given environment, the majority of the first galaxies to form 
should be
low mass systems.  Within this frame work, these low mass galaxies
merge to form more massive ones. While there is evidence for
the formation of galaxies through mergers at $z > 2$ (e.g., Conselice et al. 
2003c), there is also evidence that lower mass galaxies continue to form
their stellar populations at $z < 1$.  In fact, it appears that lower mass
galaxies finish forming after giants, through `downsizing'.  This is 
typically found for field systems in deep optical/NIR surveys 
(e.g., Cowie et al. 1996; Kauffmann et al. 2003; 
Bundy, Ellis, Conselice 2005), although convincing evidence also exits
showing that dwarf galaxies in clusters have younger light weighted stellar
population ages than giant galaxies 
(Caldwell et al. 2003; De Lucia et al. 2004) suggesting a similar, 
although likely earlier, process occurring in dense environments. 

Our results suggest that the globular clusters we observe in VCC~1386 are
old and metal poor, and thus do not fit the downsizing picture.  The reason
is that because they are old and metal poor,
they potentially formed earlier or at a similar time as stars in giant ellipticals.   What this implies is that although downsizing is a real effect, such
that some fraction of the stars in low mass galaxies form after those in
giants, this does not imply that all the stars in low mass galaxies
formed after most of the stars massive galaxies formed. 
A similar trend is seen for the 
Virgo dwarf VCC~1087, whose average metal indices are published in 
Strader et al. (2005).   VCC~1087's globular clusters have
an average, and weighted standard deviation from the mean,
values of H$\beta$=2.35$\pm$0.06, $<$Fe$>$ = 1.17$\pm$0.09 and
Mg $b$ = 0.91$\pm$0.07 (plotted on Figure~4). Globular clusters in
the Fornax dwarf elliptical surrounding
the Milky Way also has similar indices (Strader et al. 2003) which are also
plotted on Figure~4. This suggests that the globular clusters in dwarf
ellipticals are very old ($> 5$ Gyr) in both the Local Group and
the Virgo Cluster, two vastly different environments.  This interpretation
may not however be unique for globular clusters around dwarf
ellipticals as Puzia et al. (2000) find metal 
rich globulars surrounding the dwarf NGC 3115 DW1.

Why do the integrated stellar populations in the main body of dwarf ellipticals
appear to be younger than giant ellipticals, and younger than 
their own globular
clusters?  One possibility is that stellar 
population analyses are luminosity weighted, and are therefore sensitive to
any recent star formation (Trager et al. 2000).   If an equal amount of
star formation were to occur in a giant and a dwarf, it would be much
easier to identify these new stars in the lower mass system.  We know
that dwarfs in the Local group, such as the Fornax dwarf elliptical, 
have multiple and extended star formation episodes (Buonanno et al. 1999), 
including old stars 
(Grebel \& Gallagher 2004).   It is currently impossible to rule out any recent
star formation in giant ellipticals that may be the counter part of the younger
generations of stars that form in dwarfs after their globular clusters. 
GALEX observations however
suggest that star formation does occur in elliptical galaxies at low redshift 
(Yi et al. 2005), yet
the evidence for this in Virgo is currently ambiguous (Boselli et al. 2005).

If we assume that the star formation history of dwarfs in Virgo is more 
extended than
that of the giants, the obvious question is why it continues in low-mass
galaxies, but ends earlier in high mass systems. There 
are several possible explanations for this.  One is that feedback through
supernova heating is more efficient in shallower potentials
(and thus lower mass galaxies).  When stars first formed in a dwarf halo, 
supernovae and stellar winds would
have been very efficient at removing and heating gas.   This feedback 
effectively slows down the
star formation process compared to the giant galaxies.   Later, if this gas 
still resided in the halo, it would cool and form stars.  Alternatively,
AGN feedback is more effective in massive galaxies where massive black holes 
exist.  This AGN feedback would deposit enough energy to the point where 
it could halt star
formation fairly quickly (e.g., Granato et al. 2004).    It is not clear
however if 
low mass dwarf galaxies contain black holes, or go through an AGN phase,
and thus it is possible that a corresponding processes in dwarfs would not 
be as effective.

Delayed star formation in dwarf ellipticals induced by SN feedback however
presents another problem, namely why so few globulars
formed during later stages of star formation.   With the possible exception of
globular cluster \#4, it appears that all of the globulars in our sample 
are older
and more metal poor than VCC~1386 itself.  The colors of the globular
clusters surrounding VCC~1386, even those for which we do not have measured
indices, are furthermore
bluer than the light from VCC~1386 at the same projected position.  This
is also seen for the dozens of other globular cluster systems studied in the 
Virgo and Fornax clusters by the Hubble Space Telescope 
(Lotz et al. 2004).    A possible explanation is
that the Virgo cluster formed the bulk of its mass
between the time the first globulars in VCC~1386
formed, and later star formation that formed the remainder of VCC~1386.  
Due to tides, the more massive cluster could  have induced cluster 
evaporation in the lower dwarf potentials.    
The delay in star formation can also be accounted for
by the reionization of the universe (e.g., Santos 2003) 
which appears to have occurred between $z \sim 6 - 10$. 
This would allow globulars to form before reionization, and any remaining 
gas inside
these halos would photo-evaporate.   Later episodes of star formation
would then occur once the gas inside these lower mass halos cools, which
likely occurred at $z < 1$.  Cold gas
is seen in roughly $\sim$ 15\% of dwarf ellipticals in the Virgo cluster
(Conselice et al. 2003b), making this a possible scenario.  

Finally, our results are consistent with some of 
the globular clusters surrounding giant ellipticals originating from 
dwarf ellipticals.
 Both color distributions and Lick indices for globular clusters
surrounding giant galaxies
in the Virgo cluster overlap the same values found for the globulars
in VCC~1386 (e.g., Kundu et al. 1999; 
Cohen et al. 2002).     For example, giant Virgo 
elliptical galaxies, such as M87,
have blue color distributions that overlap the color distribution of our dE 
globular clusters (e.g., Kundu et al. 1999; Forbes et al. 2004; Figure~5), 
suggesting  that giant elliptical globular clusters
may have formed partially from the accretion and mergers of 
lower mass systems. However, it is impossible for
globular clusters in dwarfs to form all the globulars around  
giant galaxies, as there is also a significant red globular cluster population 
surrounding giant ellipticals (Zepf \& Ashman 1993). 
A speculative explanation for this is that that some of
these red globulars form out of cold gas transported by an accreted
dwarf, the same material the body of the dwarfs we see today were formed.  

More observations of globular clusters in dwarf elliptical galaxies in rich
environments, such as in Virgo and Fornax, are needed to determine the 
universality of these results, and whether multiple populations of globular 
clusters with differing ages and metallicities exist within dwarfs.  However,
due to the faintness of these globulars, a comprehensive 
study likely must await the advent of 20-30 meter sized telescopes.

I thank Jason Rhodes for help with the observations, Arunav Kundu for 
making his list of M87 globular cluster photometry available electronically,
and Jean Brodie and James Taylor for useful discussions.    I also acknowledge 
support from an NSF Astronomy and Astrophysics Fellowship.  I also wish
to recognize the highly significant cultural role and reverence 
that the summit of Mauna Kea has within the indigenous Hawaiian 
community, it is a privilege to be given the opportunity to conduct 
observations from this mountain.

\newpage

\begin{inlinefigure}
\begin{center}
\vspace{2cm}
\hspace{0.5cm}
\rotatebox{0}{
\resizebox{\textwidth}{!}{\includegraphics[bb = 25 25 625 625]{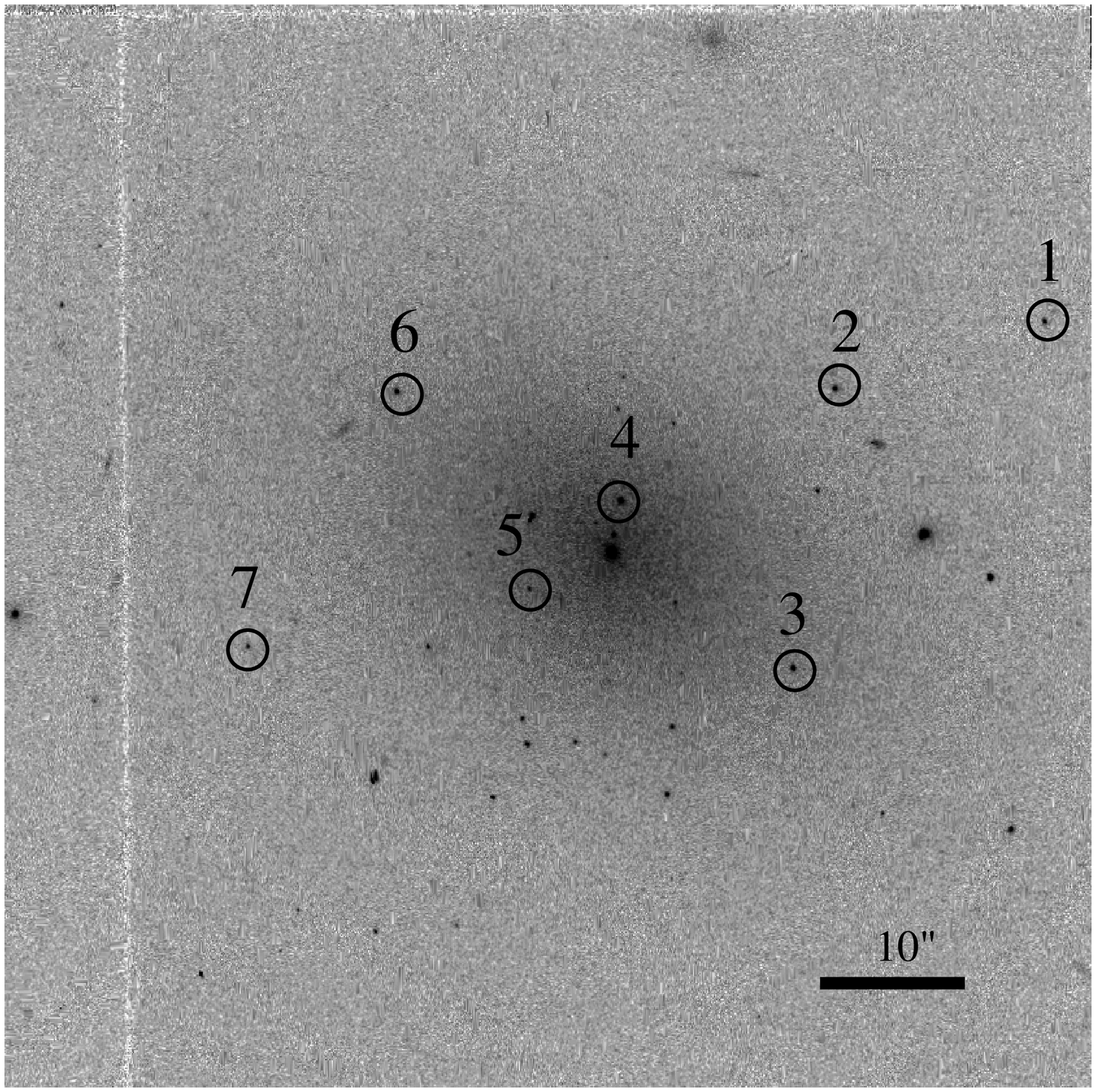}}}
\end{center}
\vspace{-2cm}
\figcaption{Hubble Space Telescope image of VCC~1386 with its globular clusters
circled and numbed.  The confirmed globular cluster members
are objects \#1,2,3,4,6. }
\end{inlinefigure}

\begin{inlinefigure}
\begin{center}
\vspace{2cm}
\hspace{-3cm}
\rotatebox{0}{
\resizebox{\textwidth}{!}{\includegraphics[bb = 25 25 525 525]{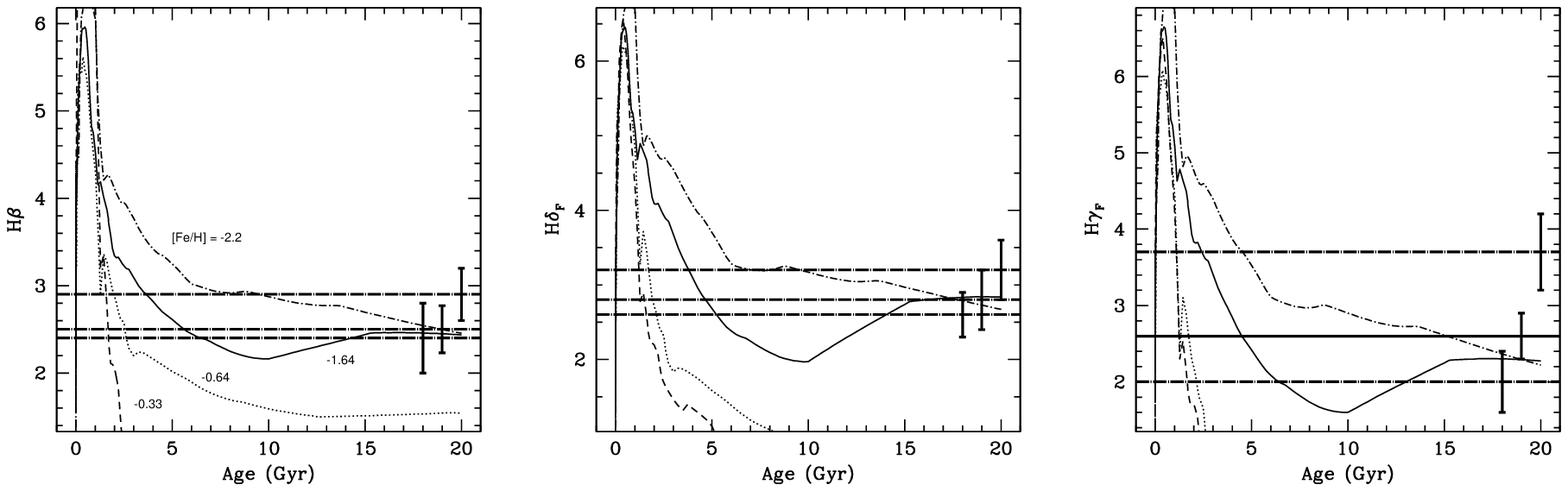}}}
\end{center}
\vspace{-9cm}
\figcaption{A comparison between the Balmer indices for our globular clusters
(H$\beta$, H$\delta_{\rm F}$ and
H$\gamma_{\rm F}$ plotted as horizontal lines) and models of how these indices evolve as a function of
time from Bruzual \& Charlot (2003).  Four models are plotted at metallicities [Fe/H] = $-2.2$, $-1.64$, $-0.64$
and $-0.33$, from top to bottom, respectively. Error bars for these
indices are plotted on the right hand side at arbitrary chosen ages
of 20, 19 and 18 Gyr. }
\end{inlinefigure}

\begin{inlinefigure}
\begin{center}
\vspace{2cm}
\hspace{-0.5cm}
\rotatebox{0}{
\resizebox{\textwidth}{!}{\includegraphics[bb = 25 25 625 625]{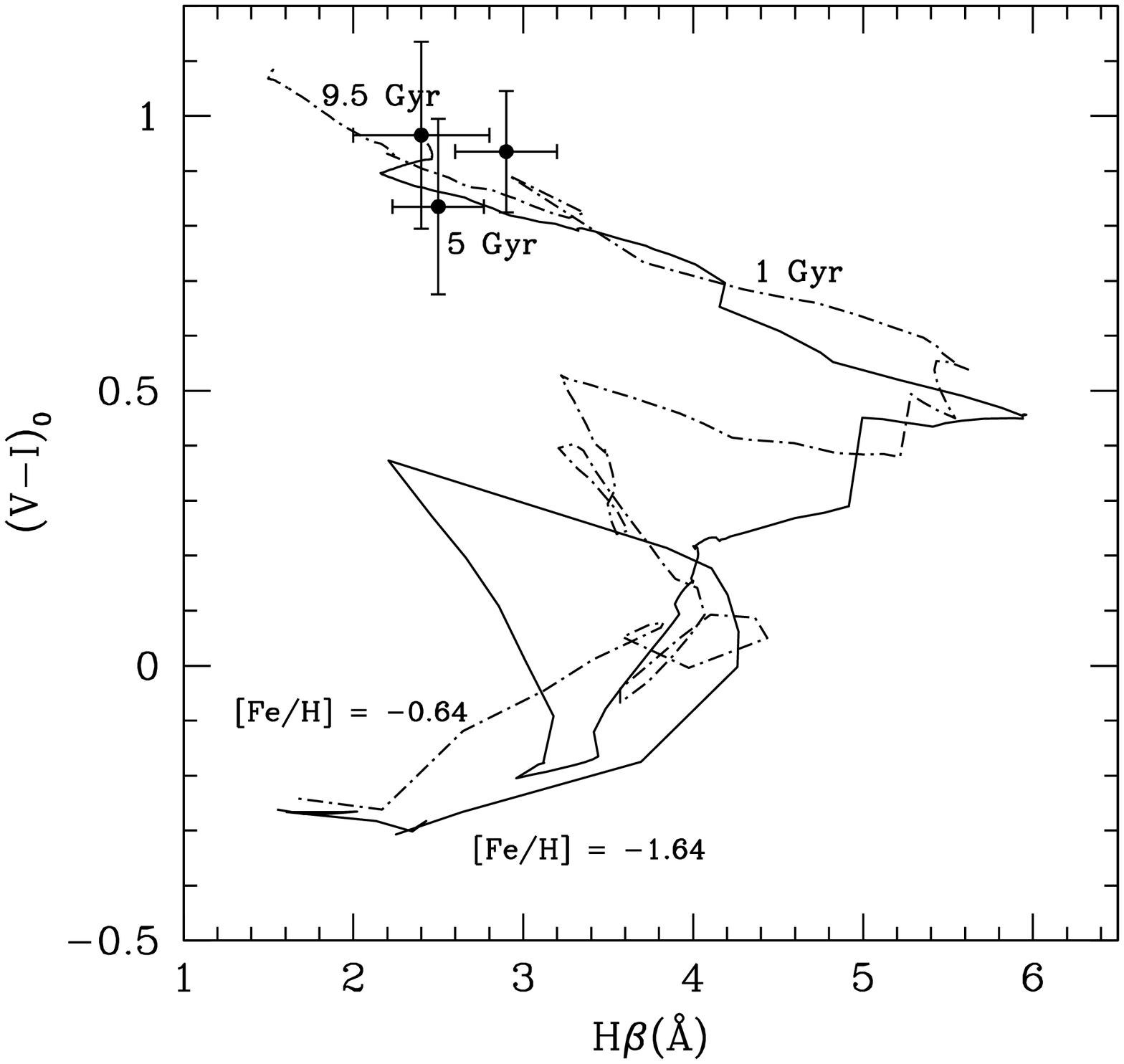}}}
\end{center}
\vspace{-4cm}
\figcaption{Plot of (V$-$I) colors and the H$\beta$ index for VCC~1386's
globular clusters.  Two Bruzual \& Charlot (2003) evolutionary models at 
metallicities
[Fe/H] = $-1.64$ and $-0.64$ are over plotted.  Other metallicities, such as
[Fe/H] = $-2.2$ and [Fe/H] = $-0.33$ show a similar evolution.
The three points with error bars represent the
values for our globular clusters.  The average position on this diagram 
where modeled
ages reach 1 Gyr, 5 Gyr and 9.5 Gyr are labeled.     }
\end{inlinefigure}

\begin{inlinefigure}
\begin{center}
\vspace{2cm}
\hspace{-2.5cm}
\rotatebox{0}{
\resizebox{\textwidth}{!}{\includegraphics[bb = 25 25 525 525]{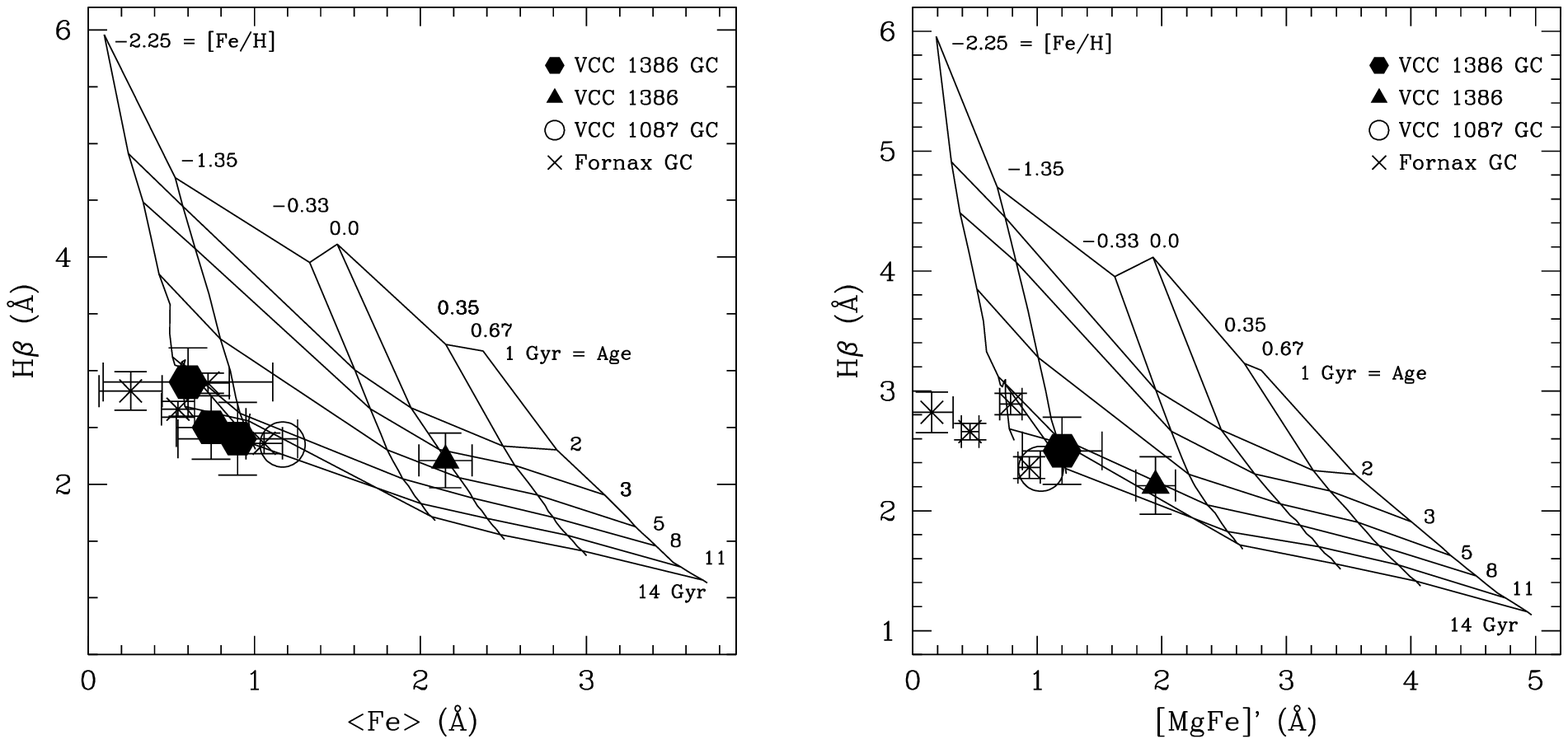}}}
\end{center}
\vspace{-8cm}
\figcaption{(a) Plot of the indices H$\beta$ and
$<$Fe$>$ for the VCC~1386 globular clusters (hexagons)
superimposed on Thomas et al. (2003) model isochrones and
isometallicity lines for $[\alpha$/Fe] = 0.3. The different ages
and metallicities along this grid are labeled.  Indices for the body
of VCC~1386 is shown as a triangle (Geha et al. 2003). Values for
the globular clusters surrounding the Local Group dwarf elliptical Fornax
are plotted as crosses (Strader et al. 2003), 
while the average values for the globular clusters surrounding the 
Virgo dE VCC~1087 is plotted as an open circle (Strader et al. 2005). 
(b) Similar to (a) except
the index [MgFe]' (see text) is plotted with H$\beta$.  }
\end{inlinefigure}

\begin{inlinefigure}
\begin{center}
\vspace{2cm}
\hspace{-0.5cm}
\rotatebox{0}{
\resizebox{\textwidth}{!}{\includegraphics[bb = 25 25 625 625]{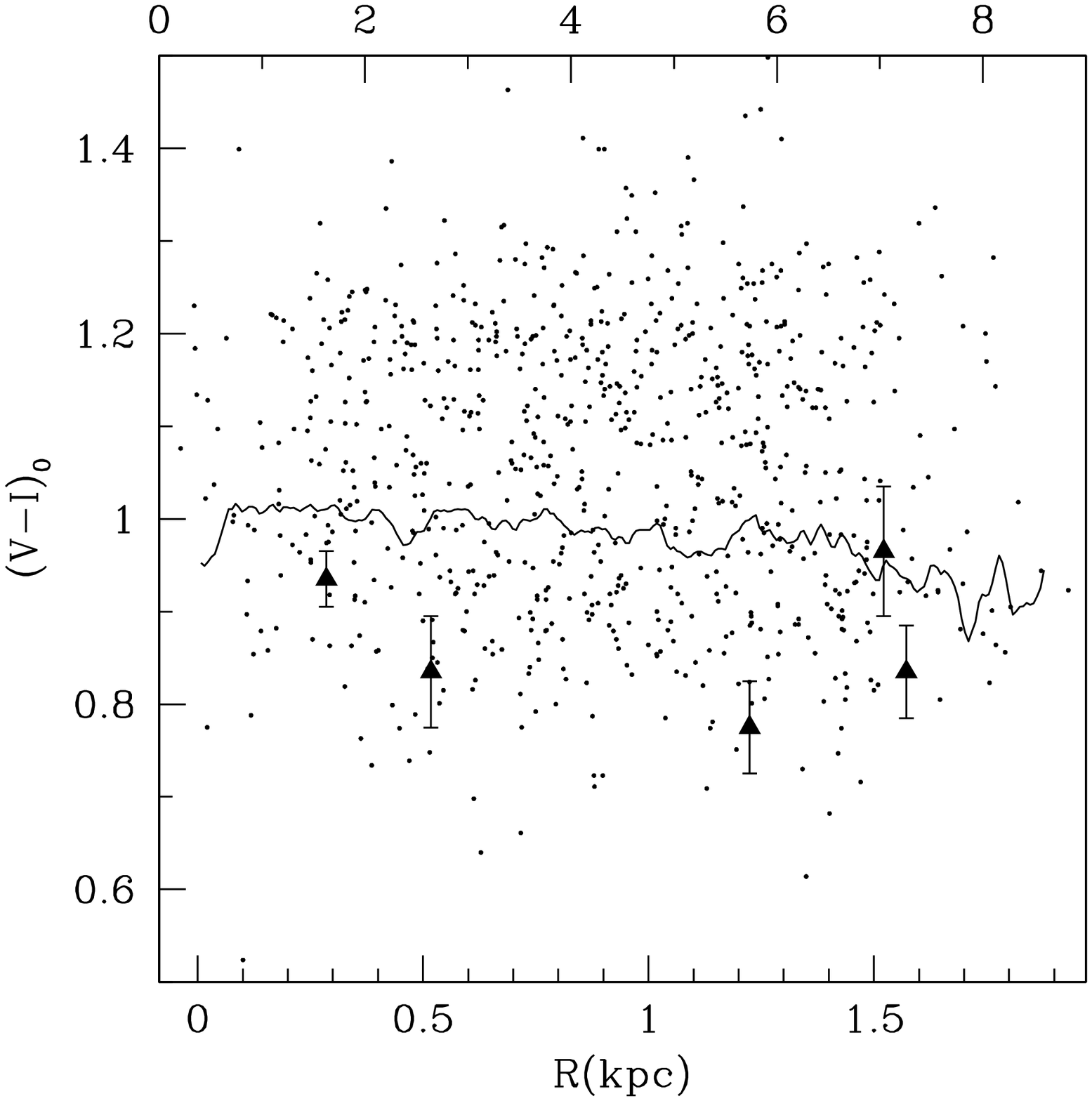}}}
\end{center}
\vspace{-4cm}
\figcaption{The radial distribution of VCC~1386 globular
cluster colors (triangles).  The solid line is the
color profile for VCC~1386. The small points and top axis give 
the radial distribution of colors for $\sim 800$ inner
globular clusters surrounding M87 out to $\sim 8$ kpc (Kundu et al. 1999). }
\end{inlinefigure}

\begin{inlinefigure}
\begin{center}
\vspace{2cm}
\hspace{-0.5cm}
\rotatebox{0}{
\resizebox{\textwidth}{!}{\includegraphics[bb = 25 25 625 625]{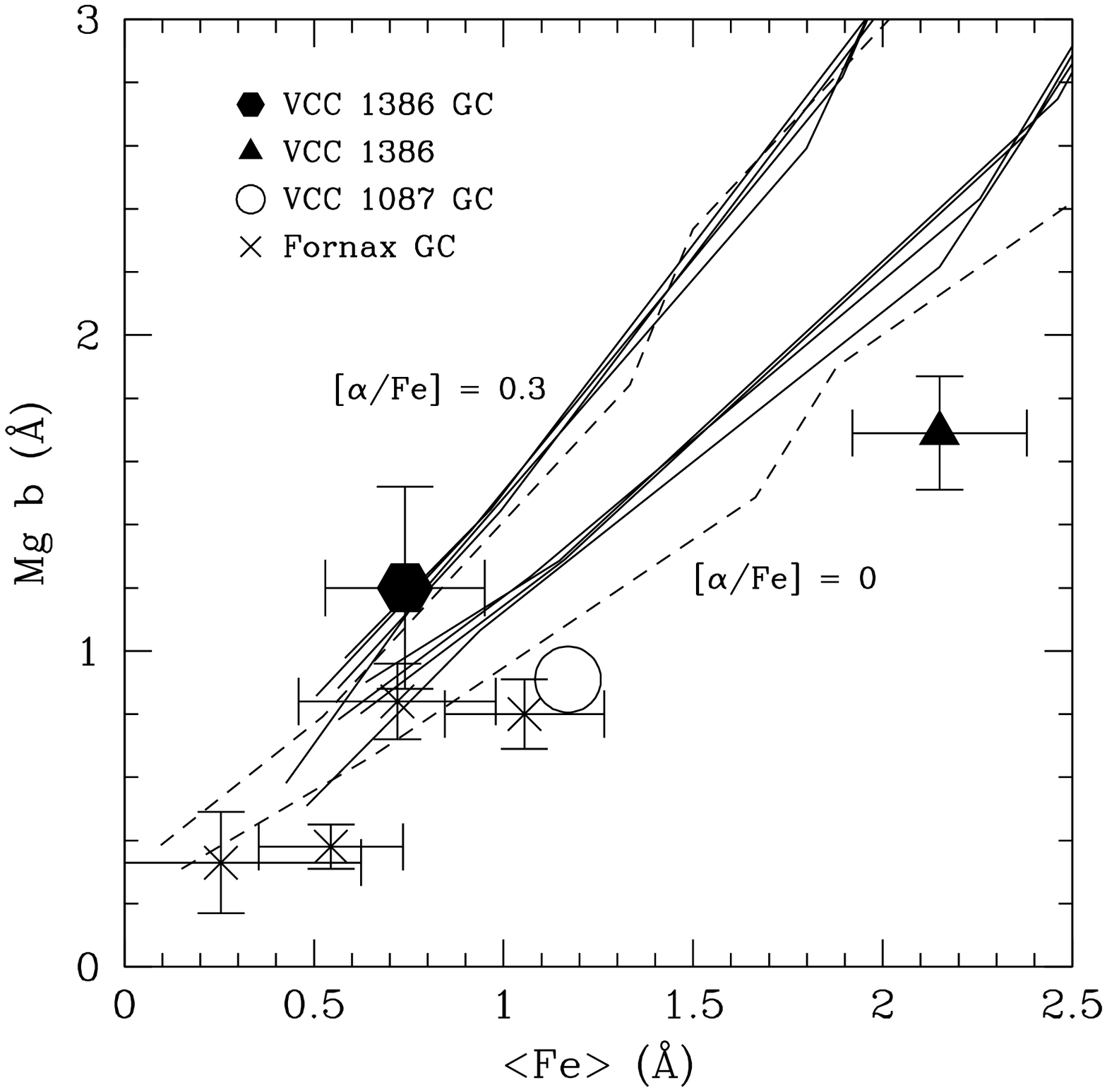}}}
\end{center}
\vspace{-4cm}
\figcaption{Figure showing the relationship between the $<$Fe$>$ index and the
Mg $b$ index for globular cluster \#2.  Also plotted are the same indices 
for globulars found in
the Fornax dE, the average values for globulars in 
another Virgo dwarf elliptical,
VCC~1087 (Strader et al. 2005), and the value for the body of VCC~1386.  
Models are over-plotted as lines showing the relationship between $<$Fe$>$ and 
Mg $b$ at two different values of $\alpha$ ($\alpha$=0.3 and $\alpha$=0). 
The individual model
lines show this relationship at ages 1,5,8,11 and 14 Gyr, where the 
youngest
age, 1 Gyr, is plotted as the dashed line at each value at each $\alpha$ 
value.}
\end{inlinefigure}

\begin{inlinefigure}
\begin{center}
\vspace{2cm}
\hspace{-3cm}
\rotatebox{0}{
\resizebox{\textwidth}{!}{\includegraphics[bb = 25 25 525 525]{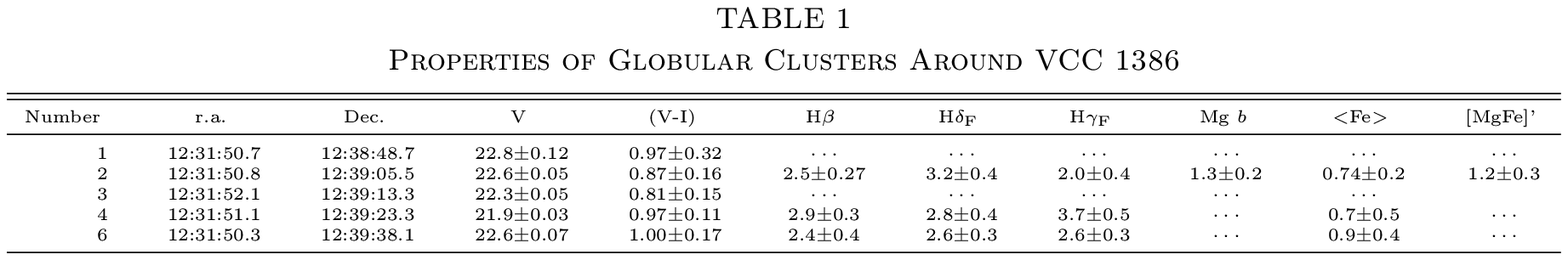}}}
\end{center}
\vspace{-2cm}
\end{inlinefigure}

\end{document}